\documentclass[aps,pra,reprint,groupedaddress,showkeys]{revtex4-1}
\usepackage{amsmath,graphicx,verbatimbox,array,float}
\usepackage{indentfirst}
\usepackage{braket}
\usepackage{diagbox}
\usepackage{amssymb}
\usepackage{bbold}
\usepackage{colortbl}
\usepackage{multirow}
\usepackage{subcaption}
\newcommand{\figref}[1]{Fig.~\ref{#1}}

\renewcommand{\eqref}[1]{Eq.~(\ref{#1})}
\newcommand{\tr}{{\mathrm{Tr}}}

\begin{document}
\title{Efficiency of photonic state tomography affected by fiber attenuation}
\author{Artur Czerwinski}\email[]{aczerwin@umk.pl} 
\author{Jakub Szlachetka}
\affiliation{Institute of Physics, Faculty of Physics, Astronomy and Informatics \\ Nicolaus Copernicus University in Torun, ul. Grudziadzka 5, 87--100 Torun, Poland}

\begin{abstract}
In this article, we investigate the efficiency of photonic state tomography in the presence of fiber attenuation. The theoretical formalism of the photon loss is provided by implementing methods from the theory of open quantum systems. The quantum state is reconstructed from photon counts obtained for symmetric informationally complete POVMs. The number of photons that reach the detectors is numerically modeled by the binomial distribution, which describes the loss of light caused by the medium. This approach allows us to study the quality of state tomography versus the length of the fiber. In particular, we focus on entangled qubits and qutrits, which are sent through fibers of different lengths. The amount of entanglement detected by the measurement scheme is quantified and presented on graphs. The results demonstrate how the quality of photonic tomography depends on the distance between the source and the receiver.
\end{abstract}
\maketitle

\section{Introduction}

Quantum key distribution (QKD) relies on exchanging a secure key between two distant parties through a non-trusted communication channel \cite{Pirandola2020}. The most common implementation of QKD is based on using single photons as a carrier of a quantum cryptographic key \cite{BB84,BB84v2,Bennett1992}. In other protocols, the security of the QKD process is guaranteed by the entanglement of the state used to encode the key \cite{Ekert1991,Waks2002}. QKD systems are built on the existing communication channels utilized in modern telecommunication to increase the utility of such systems. The most versatile channel for transmitting information is the optical fiber network. Losses in an optical fiber increase exponentially with the length $L$ of the channel used and the attenuation coefficient denoted by $\alpha$. The value of the attenuation factor, usually expressed in decibels per kilometer (dB/km), depends on the fiber material and the manufacturing tolerances, but it also varies with the wavelength. Fiber optic losses are mainly due to material absorption. When the distance to exchange cryptographic keys is above $300$ km, the losses become significant \cite{Yin2016,Chen2020,Fang2020}. This is related to the inability to duplicate the entangled states used to exchange the cryptographic key because of the no-cloning theorem \cite{Wotters1982}.

To investigate the impact of fiber attenuation on photon transmission, we implement quantum state tomography (QST), which allows us to obtain the density matrix representing the quantum state from measurable data \cite{Paris2004}. Methods of QST are commonly utilized to reconstruct the state of photons produced by a source \cite{White1999,Horn2013}. Different characteristics of a photonic state can be determined if its density matrix is reconstructed. More specifically, for two-photon states, one can quantify the entanglement by using a variety of entanglement measures \cite{Horodecki2001,Eisert2022}.

In this paper, we present a comprehensive framework of photonic state reconstruction and entanglement quantification that takes into account fiber attenuation. First, we introduce theoretical formalism by characterizing the photon loss as a non-unitary decoherence process that affects the Fock state. We prove that the decline in the number of photons can be represented by either  Kraus operators or a quantum generator. As a consequence, the Fock state is subject to a legitimate quantum evolution in the domain of fiber length.

The formalism allows us to investigate quantum tomography of states encoded in a photon's degree of freedom. We assume that the number of photons that successfully pass through the fiber is represented by a binomial distribution. Then, the shot noise related to photon-counting is imposed to make the scheme realistic. The efficiency of state reconstruction is studied in various settings. The fidelity of quantum states is used as an indicator of the precision of state recovery. As for entanglement quantification, we implement two measures -- the concurrence, which relates to two-qubit states, and the negativity, which works well for two-qutrit states.

Throughout the article, we follow the bra-ket notation to denote pure quantum states. We operate in finite-dimensional spaces with standard bases, which allows us to represent the density operator as a matrix. To avoid ambiguity, the Fock state representing the number of photons in a beam is denoted by $\varrho$, whereas a quantum state encoded in a degree of freedom is given by $\rho$.

In Sec. II, we introduce the theoretical formalism of the photon loss. The framework of state tomography is presented in Sec. III, starting from the measurement scheme and noise model. Then, entanglement measures are defined. Next, in Sec. IV, we present and discuss the results devoted to qubit tomography. Finally, in Sec. V, we analyze the figures concerning qutrits. The findings of the article provide valuable insight into the impact of fiber attenuation on state tomography and entanglement detection. The work is concluded in the final section, where we also indicate problems for future research.

\section{Theoretical model of the photon loss in fiber transmission}\label{theorframework}

According to the Beer-Lambert law, if a source generates a beam with the initial power $P_0$ that is transmitted through a fiber of the length $L$, the receiver gets the output power that can be expressed as
\begin{equation}\label{eq3}
    P_{out} (L) = P_0 \: 10^{- \frac{\alpha L}{10}},
\end{equation}
where $\alpha$ stands for the attenuation coefficient that characterizes the fiber. Conventionally, $\alpha$ is given in dB/km. For simplicity, the law \eqref{eq3} can be expressed as $P_{out} (L) = P_0 \: e^{-\Lambda t}$, where $\Lambda \equiv \ln 10 \:\alpha /10$.

If we implement this law in the single-photon framework, it implies that one photon can successfully pass through the fiber with probability $e^{- \Lambda L}$ whereas $1-e^{- \Lambda L}$ gives the probability of a failure (photon loss). Assuming that the source generates $\mathcal{N}$ photons in beam, the number of particles  that reach the receiver (denoted by $\widetilde{\mathcal{N}}$) can be modeled by the binomial distribution, i.e., $\widetilde{\mathcal{N}} \in \mathcal{B} (\mathcal{N}, e^{- \Lambda L})$. By applying the binomial distribution, we can study the photon loss as a decoherence process affecting the Fock state of the beam. We consider two scenarios. First, we introduce the formalism for one photon traveling through the fiber, and then we generalize it for a beam consisting of $\mathcal{N}$ photons.

\subsection{Decoherence of a one-photon state}

One photon traveling through a fiber can be considered a two-level system associated with an orthonormal basis $\{\ket{0},\ket{1}\}$. The vectors correspond to physical situations: ``There is a photon" (vector $\ket{1}$) and ``There is no photon" (vector $\ket{0}$). The initial state is given by $\varrho_{in} = \ket{1}\!\bra{1}$ since the source is assumed to emit a photon with certainty. By imposing the binomial distribution, we can describe the evolution of this state versus the fiber length:
\begin{equation}\label{fock1}
    \varrho (L) = \left(1 -  e^{- \Lambda L} \right)\ket{0}\!\bra{0} +  e^{- \Lambda L} \ket{1}\!\bra{1},
\end{equation}
which can be expressed, equivalently, as:
\begin{equation}\label{op1}
\varrho (L) =  K_0 (L) \,\varrho_{in}\, K_0^{\dagger}(L) + K_1 (L) \,\varrho_{in}\, K_1^{\dagger}(L),
\end{equation}
where
\begin{equation}
K_0 (L) =  \begin{pmatrix} 0 &  \sqrt{1- e^{- \Lambda L}} \\ 0  & 0 \end{pmatrix},\hspace{0.05cm} K_1 (L) = \begin{pmatrix} 1 &  0 \\ 0  & \sqrt{e^{- \Lambda L}} \end{pmatrix}.
\end{equation}
From \eqref{op1}, we see that the process of photon loss is a type of non-unitary decoherence that can be described by a Kraus representation \cite{Sudarshan1961,Kraus1983}. Furthermore, one can notice that for any $L\geq0$, we have $K_0^{\dagger}(L) K_0 (L)+K_1^{\dagger}(L) K_1 (L)=\mathbb{I}_2$ (where by $\mathbb{I}_d$ we denote a $d\times d$ identity matrix), which implies that the map preserves the trace of the density matrix. As a result, the operation \eqref{op1} provides a one-parameter continuous family of complete positive and trace-preserving (CPTP) maps. This implies that the photon loss can be considered a legitimate quantum dynamics of the Fock state in the domain of fiber length.

The formalism can be further developed by differentiating \eqref{op1}, which results in
\begin{equation}\label{op2}
    \frac{d \varrho (L)}{d L} = \Lambda \left(E_{01} \varrho(L) E_{01}^{\dagger} - \frac{1}{2} \left\{E_{01}^{\dagger}E_{01},  \varrho(L) \right\}  \right),
\end{equation}
where $E_{01} =\ket{0}\!\bra{1}$ and $\{X, Y\}$ denotes the anticommutator, i.e., $\{X, Y\} = XY + YX$. The right-hand side of \eqref{op2} is a specific example of Gorini-Kossakowski-Sudarshan-Lindblad (GKSL) quantum generator \cite{Gorini1976,Lindblad1976}. $E_{01}$, which describes the decay from $\ket{1}$ to $\ket{0}$, can be called a ``jump operator".

The framework of the photon loss is in line with the theory of open quantum systems. In particular, when we consider a one-photon Fock state, we notice a direct analogy to the amplitude damping channel \cite{Nielsen2000}. Such a quantum channel applied to a two-level atom can describe, for example, spontaneous emission, which involves a decay from a higher energy level to the ground state, cf. \cite{Weisskopf1930,Sharma1990}.

The theory of one-photon decoherence can be considered in the context of randomness in quantum mechanics \cite{Bera2017}. The concept of random numbers is an important question for many disciplines. However, generation of high-quality randomness has been a difficult problem \cite{Neumann1952,Markowsky2014}. According to the physics-based approach, random numbers should be unpredictable to any observer who is constrained by the laws of physics \cite{Pironio2010,Acin2016}. In our model, we can obtain a quantum random number generator (QRNG) since the attenuation process, which is inherently uncontrollable and unpredictable, leads to the realization of perfectly random bits. For any attenuation coefficient $\alpha$, we assume to be able to adjust the fiber length according to
\begin{equation}\label{qrng1}
    L (\alpha) = \frac{10 \,\ln 2}{ \ln 10} \;\frac{1}{\alpha},
\end{equation}
which guarantees that the one-photon Fock state \eqref{fock1} takes the form of the maximally mixed state. Consequently, if we consider a sequence of $M$ photons separated in time, we get the state
\begin{equation}\label{randoms}
    \varrho_{rand} = \left(\frac{1}{2}\ket{0}\!\bra{0}  +\frac{1}{2} \ket{1}\!\bra{1}  \right)^{\otimes M}.
\end{equation}
The state $\varrho_{rand}$ represents $M$ realizations of a perfect random bit, where the values $0$ and $1$ are taken with equal probability. For each time slot, if the detector clicks to announce a photon, we obtain $1$. Otherwise, we have $0$.

\begin{figure}[h]
\centering
\includegraphics[width=\linewidth]{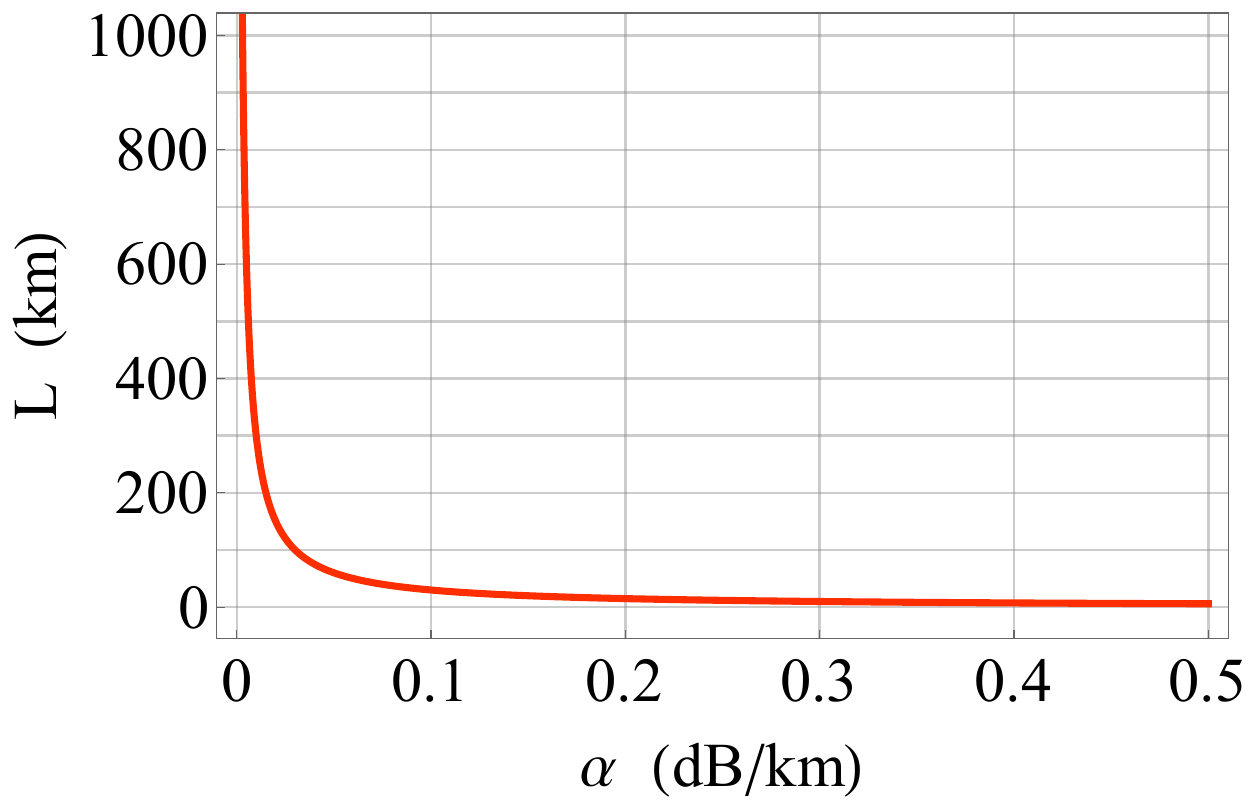}
\caption{The fiber length required for quantum randomness versus attenuation.}
\label{qrng2}
\end{figure}

Furthermore, the relation between $L$ and $\alpha$ can be presented graphically, which is shown in \figref{qrng2}. From \eqref{qrng1}, we see that $L$ and $\alpha$ are inversely proportional. This feature appears to be a key limiting factor of this scheme of randomness generation. For a typical fiber, we have $\alpha = 0.2$ dB/km, which gives the fiber length required to achieve random states: $L \approx 15$ km. If one would like to extend the distance for transmission of random states, fibers with lower attenuation coefficients would be necessary, as presented in \figref{qrng2}.

The model presented in this work belongs to the category of intrinsic randomness, which occurs when knowledge about the initial state is not sufficient to predict future evolution. Probabilities  are used here to represent the quantum state as a statistical mixture \eqref{randoms}, which is a necessary and inevitable tool to
describe the behavior of a system subject to a randomized physical process \cite{Bera2017}.

\subsection{Decoherence of an $\mathcal{N}$-photon state}

In a general case, we consider a beam that consists of $\mathcal{N}$ photons traveling through a fiber. Then, the initial Fock state of the ensemble can be denoted by $\varrho_{in} = \ket{\mathcal{N}}\!\bra{\mathcal{N}}$. As already explained, the number of photons that pass through the medium can be modeled by a binomial distribution. As a result, the Fock state of the beam can be written as
\begin{equation}\label{op3}
     \varrho (L) = \sum_{j=0}^{\mathcal{N}} \binom{\mathcal{N}}{j} (e^{-\Lambda L})^{j} (1-e^{-\Lambda L})^{\mathcal{N}-j} \ket{j}\!\bra{j},
\end{equation}
where, for simplicity, we assume that the vectors $\{\ket{0}, \ket{1}, \dots, \ket{\mathcal{N}} \}$ form the standard basis. By convention, $\binom{\mathcal{N}}{j}$ stands for the binomial coefficient, i.e., $\binom{\mathcal{N}}{j} \equiv \mathcal{N}!/j! (\mathcal{N}-j)!$.

The spectral decomposition of the density matrix \eqref{op3} can be put into
\begin{equation}\label{op4}
    \varrho (L) = \sum_{j=0}^{\mathcal{N}}  K_{j} (L) \,\varrho_{in}\, K_{j}^{\dagger}(L),
\end{equation}
with the Kraus operators defined as
\begin{equation}\label{op5}
    K_j (L) = \begin{cases} \sqrt{P_{\mathcal{N}} (j)}\, \ket{j}\!\bra{\mathcal{N}} \:\:\text{for}\:\: j =0,1,\dots, \mathcal{N}-1 \\\\ \mathrm{diag} \left(1, \dots, 1, \sqrt{P_{\mathcal{N}} (\mathcal{N})}\right) \:\:\text{for}\:\: j =  \mathcal{N} \end{cases},
\end{equation}
where $P_{\mathcal{N}} (j) \equiv \binom{\mathcal{N}}{j} (e^{-\Lambda L})^{j} (1-e^{-\Lambda L})^{\mathcal{N}-j}$ and by $ \mathrm{diag} (1, \dots, 1, \sqrt{P_{\mathcal{N}} (\mathcal{N})})$ we denote a $(\mathcal{N}+1)\times(\mathcal{N}+1)$ diagonal matrix in which the first $\mathcal{N}$ entries of the main diagonal are all one, while the last element is $\sqrt{P_{\mathcal{N}} (\mathcal{N})}$. Furthermore, one can notice that:
\begin{equation}\label{op6}
\begin{aligned}
&    \sum_{j=0}^{\mathcal{N}}  K_{j}^{\dagger}(L) K_{j} (L) =\\& (1-P_{\mathcal{N}} (\mathcal{N})) \ket{\mathcal{N}}\!\bra{\mathcal{N}} + \mathrm{diag} \left(1, 1, \dots, P_{\mathcal{N}} (\mathcal{N})\right) = \mathbb{I}_{\mathcal{N}+1},
\end{aligned}
\end{equation}
which proves that the map \eqref{op4} is CPTP for any $L\geq 0$. This demonstrates that the transmission of an $\mathcal{N}$-photon state can be described by a legitimate model of non-unitary decoherence.

The formalism can be developed by differentiating \eqref{op3}, which leads to a master equation:
\begin{widetext}
\begin{equation}\label{op7}
\begin{aligned}
 \frac{d \varrho (L)}{d L} =& \sum_{j=0}^{\mathcal{N}} \binom{\mathcal{N}}{j}  \left( - j \Lambda (e^{-\Lambda L})^{j} (1-e^{-\Lambda L})^{\mathcal{N}-j} + (\mathcal{N}-j) \Lambda  e^{-\Lambda L}  (e^{-\Lambda L})^{j} (1-e^{-\Lambda L})^{\mathcal{N}-j -1}   \right) \ket{j}\!\bra{j} =  \\&
 \Lambda \left[- \sum_{j=1}^{\mathcal{N}} \binom{\mathcal{N}}{j} j (e^{- L})^{j} (1-e^{-\Lambda L})^{\mathcal{N}-j} + \sum_{j=0}^{\mathcal{N}-1} \binom{\mathcal{N}}{j+1} (j+1) (e^{- L})^{j+1} (1-e^{-\Lambda L})^{\mathcal{N}-(j+1)}  \right]\ket{j}\!\bra{j} =\\&
 \Lambda \sum_{j=1}^{\mathcal{N}} j \left[  \binom{\mathcal{N}}{j} (e^{- L})^{j} (1-e^{-\Lambda L})^{\mathcal{N}-j}  \ket{j-1}\!\bra{j-1}  - \binom{\mathcal{N}}{j} (e^{- L})^{j} (1-e^{-\Lambda L})^{\mathcal{N}-j} \ket{j}\!\bra{j} \right]=\\&
 \sum_{j=1}^{\mathcal{N}} \Lambda\, j \left( E_{(j-1)j} \varrho (L) E_{(j-1)j}^{\dagger} - \frac{1}{2} \left\{ E_{(j-1)j}^{\dagger}E_{(j-1)j}, \varrho (L)  \right\} \right),
\end{aligned}    
\end{equation}
\end{widetext}
where $E_{(j-1)j} = \ket{j-1}\!\bra{j}$ represents the jump operator from the $j-$th to $(j-1)-$th state (for $j=1,\dots, \mathcal{N}$). The last line of \eqref{op7} shows that the process of attenuation of an $\mathcal{N}-$photon state can also be analyzed within the master equation approach by assigning a GKSL generator.

Moreover, the spectral decomposition \eqref{op3} enables one to study the purity:
\begin{equation}
    \gamma (L) := \tr \varrho^2(L) = \sum_{j=0}^{\mathcal{N}} P_{\mathcal{N}}^2 (j)
\end{equation}
and the von Neumann entropy:
\begin{equation}
    S (L) := - \sum_{j=0}^{\mathcal{N}} P_{\mathcal{N}} (j) \ln P_{\mathcal{N}} (j)
\end{equation}
of the Fock state $\varrho (L)$ subject to fiber attenuation.

\begin{figure}[h]
\centering
\includegraphics[width=\linewidth]{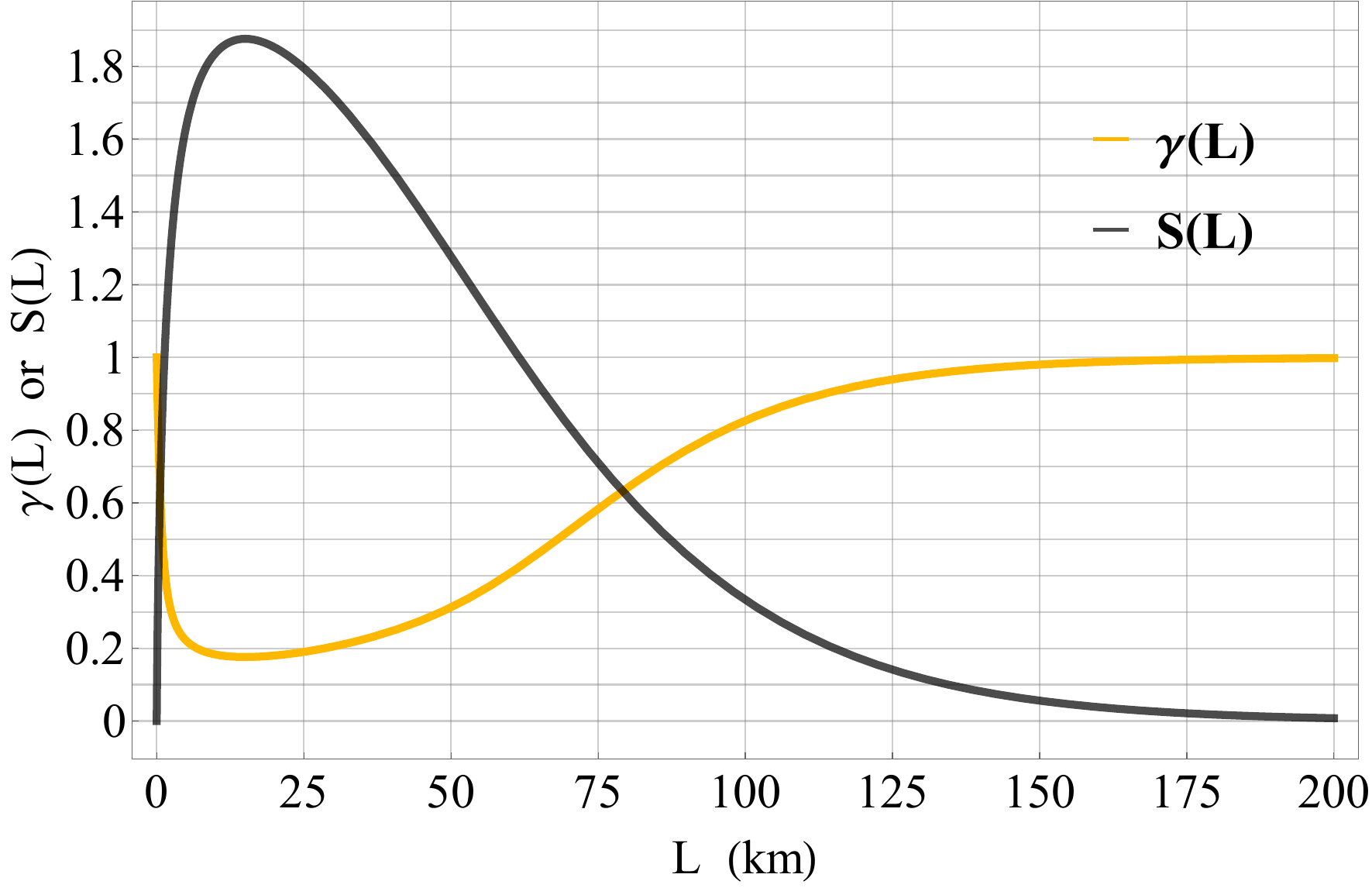}
\caption{The von Neumann entropy and purity of a Fock state associated with a beam traveling through a fiber such that $\alpha = 0.2$ dB/km. Initially, the beam consisted of $10$ photons.}
\label{purity-entropy}
\end{figure}

 In \figref{purity-entropy}, one finds the plots of both quantities for a beam comprising $10$ photons that travel through a fiber with the attenuation coefficient $\alpha = 0.2$ dB/km. The plots reflect the structure of the generator of evolution given in \eqref{op7} because the initial Fock state $\varrho_{in} = \ket{10}\!\bra{10}$ is quickly deprived of its coherence, which is shown as a boost in entropy and a decline in purity. Then, as we increase the fiber length, the state slowly converges to the final pure state, which is $\lim_{L\rightarrow \infty} \varrho (L) = \ket{0}\!\bra{0}$.

The theoretical model allows us to describe the process of attenuation within the framework of open quantum systems. In other words, the photon loss can be treated as a type of non-unitary decoherence that disturbs the Fock state representing the number of photons in the fiber.

The exponential decay of the number of photons can significantly distort measurements based on photon counting. If we encode a quantum state on a photon by exploiting a particular degree of freedom (e.g., polarization, temporal), then photon counts obtained from an experiment will depend not only on the type of measurement but also on the length of the fiber. Therefore, we propose a numerical framework to investigate the impact of fiber attenuation on quantum state tomography and entanglement quantification.

\section{Methods of state estimation and entanglement quantification}\label{methodsection}

\subsection{Quantum state tomography}

To reconstruct an unknown quantum state encoded on a photon, we implement a measurement scheme that is based on symmetric informationally complete positive operator-valued measures (SIC-POVMs) \cite{Renes2004,Fuchus2017}. For qubit tomography, the SIC-POVM involves four measurement operators, whereas, for qutrits, nine operators are required \cite{PaivaSanchez2010}. To estimate the state of entangled qubits, we construct $16$ two-qubit measurement operators by implementing the tensor product. Analogously, for entangled qutrits, we obtain $81$ operators,

In general, let us denote the measurement operators by: $M_1, \dots, M_{\eta}$, where $\eta$ indicates the necessary number of operators. Then, we assume that the source generates a beam containing $\mathcal{N}$ photons (or photon pairs) per measurement. Each photon is prepared in an identical quantum state that is described by a density matrix $\rho_{x}$. The quantum state can be encoded in a photon's degree of freedom such that one can realize the SIC-POVM scheme. In particular, one can consider polarization \cite{Rehacek2004}, spatial \cite{Pimenta2013}, or time-bin quantum states \cite{Sedziak2020}. This allows us to write a formula for the expected photon count:
\begin{equation}\label{eq1}
    e_{k} = \lceil \mathcal{N}\: \tr M_k \rho_{x} \rfloor,
\end{equation}
where the symbol $\lceil a \rfloor$ denotes rounding $a$ to the nearest integer since the number of photons cannot be fractional. The density matrix $\rho_{x}$ remains unknown to the observer and, for this reason, we follow the Cholesky decomposition to parameterize it depending on its dimension, cf. \cite{James2001,Altepeter2005}.

The formula \eqref{eq1} models the photon counts according to the Born rule, which is a theoretical foundation for this measurement scheme. However, in practice, any act of measurement involves errors and uncertainty, which implies that the values provided by the detection system will be different from what one may expect. In quantum optics, we encounter the shot noise that describes the fluctuations of the number of photons counted by the system \cite{Hasinoff2014}. As a result, the measured counts, $\{m_k\}$, are statistically independent Poissonian random variables. Thus, for an input state $\rho_{in}$, we simulate an experimental scenario by selecting measured counts randomly from a Poisson distribution: $m_k \in \mathrm{Pois} (n)$ with the mean value given by
\begin{equation}\label{eq2}
    n = \lceil \widetilde{\mathcal{N}} \: \tr M_k \rho_{in} \rfloor,
\end{equation}
where $\widetilde{\mathcal{N}}$ represents the number of photons that reached the detection system after passing through the fiber. The value of $\widetilde{\mathcal{N}}$ is generated randomly by taking into account fiber attenuation. As explained in Sec.~\ref{theorframework}, we follow the binomial distribution to simulate experimental scenario, i.e., $\widetilde{\mathcal{N}} \in \mathcal{B} (\mathcal{N}, e^{- \Lambda L})$. Similarly, for a two-photon state, we assume that the source generates $\mathcal{N}$ photon pairs per measurement, and each of them travels through a separate fiber (fiber lengths are denoted by $L_1$ and $L_2$). Since we are interested in detecting coincidences, both photons need to arrive at the corresponding detectors. Consequently, for two-photon states, the number of photon pairs that reach the detection system can be modeled by implementing the joint probability, which means: $\widetilde{\mathcal{N}} \in \mathcal{B} \left(\mathcal{N}, e^{- (\Lambda_1 L_1 + \Lambda_2 L_2)}\right)$.

The above-described approach allows one to numerically generate photon counts that correspond to a realistic scenario for any input density matrix $\rho_{in}$. Then, we follow the method of least squares (LS) to determine how well one can reconstruct the density matrix in spite of the uncertainty, cf. \cite{Acharya2019,Czerwinski2021}. This means that we search for the minimum value of the function:
\begin{equation}\label{eq4}
    f_{LS} (t_1, t_2, \dots) = \sum_{k} (e_k - m_k)^2,
\end{equation}
where $t_1, t_2, \dots$ denote the set of real parameters that characterize the density matrix $\rho_x$.

To evaluate the performance of QST in the presence of fiber attenuation, we compute, for any input density matrix $\rho_{in}$, its fidelity with the result of estimation $\rho_x$ \cite{Nielsen2000}:
\begin{equation}\label{eq5}
    F[\rho_{in}, \rho_{x}] := \left( \tr \sqrt{\sqrt{\rho_{in}} \rho_x\sqrt{\rho_{in}}} \right)^2,
\end{equation}
which is known as the Uhlmann-Jozsa fidelity that measures the closeness of two quantum states \cite{Jozsa1994,Uhlmann1976}. The efficiency of the framework depends on the properties of the input state. Therefore, to find an indicator of the average performance, we first select a sample of input states, then each of them undergoes the procedure of QST, and, finally, the average fidelity for the sample is computed. Additionally, we calculate sample standard deviation (SD) to quantify the statistical dispersion. In our study, the length of the fiber is considered an independent variable, which implies that the average fidelity can be treated as a function of $L$ and denoted by $F_{av} (L)$. As a result, this figure of merit can be plotted to observe how the efficiency of QST changes as we increase $L$. Analogously, for two-photon states, we treat the average fidelity denoted by $F_{av} (L_1, L_2)$ as a function of two variables.

\subsection{Entanglement quantification}

Apart from evaluating the precision of state tomography, we also quantify the amount of entanglement detected by the scheme. For two-qubit states, we implement the concurrence, which can be computed directly for any density matrix $\rho_x$ obtained from the scheme \cite{Hill1997,Wootters1998}. The concurrence is an entanglement monotone, which gives $C[\rho] = 0$ for a separable state $\rho$, and $C[\rho]=1$ for $\rho$ representing a maximally entangled state. This figure of merit is commonly used to quantify entanglement with imperfect measurements \cite{Walborn2006,Neves2007}.

In our application, we consider entangled photon pairs such that each photon can travel through a fiber of a different length, denoted by $L_1$ and $L_2$. Thus, for the reconstructed two-qubit state $\rho_x$, we compute the concurrence, treating the fiber lengths as independent variables. Finally, the average concurrence for a sample of input states, denoted by $C_{av} (L_1,L_2)$, can be plotted like a two-variable function. This approach allows us to observe the quantity of entanglement embraced by the reconstructed states for different combinations of fiber lengths.

As for two-qutrit states, we utilize the negativity, which is a measure of quantum entanglement that can be relatively convenient for an arbitrary bipartite system \cite{Vidal2002}. For a $9\times 9$ density matrix $\rho_{x}$ obtained from the scheme, we compute
\begin{equation}\label{eq6}
    \mathrm{N} [\rho_{x}] = \frac{\|\rho_{x}^{\Gamma_A}\|_1-1}{2},
\end{equation}
where $\rho_{x}^{\Gamma_A}$ denotes the partial transpose of $\rho_{x}$ with respect to the subsystem $A$ and $\|\sigma\|_1$ represents the trace norm of $\sigma$, i.e., $\|\sigma\|_1 := \tr \sqrt{\sigma^{\dagger}\sigma}$. The formula \eqref{eq6} can be implemented numerically to allow for straightforward entanglement quantification. It is worth stressing that negativity is an entanglement monotone, but it does not always guarantee entanglement detection since, for PPT entangled states, it results in zero. However, in our framework, we operate with a sample of maximally entangled qutrits, which implies that negativity can be considered a proper entanglement measure. In the same vein as with the concurrence, we compute the average negativity for a sample and plot it as a two-variable function, denoted by $\mathrm{N}_{av} (L_1, L_2)$.

\section{Qubit tomography}

\subsection{Single qubits}

\begin{figure}[h]
\centering
\includegraphics[width=\linewidth]{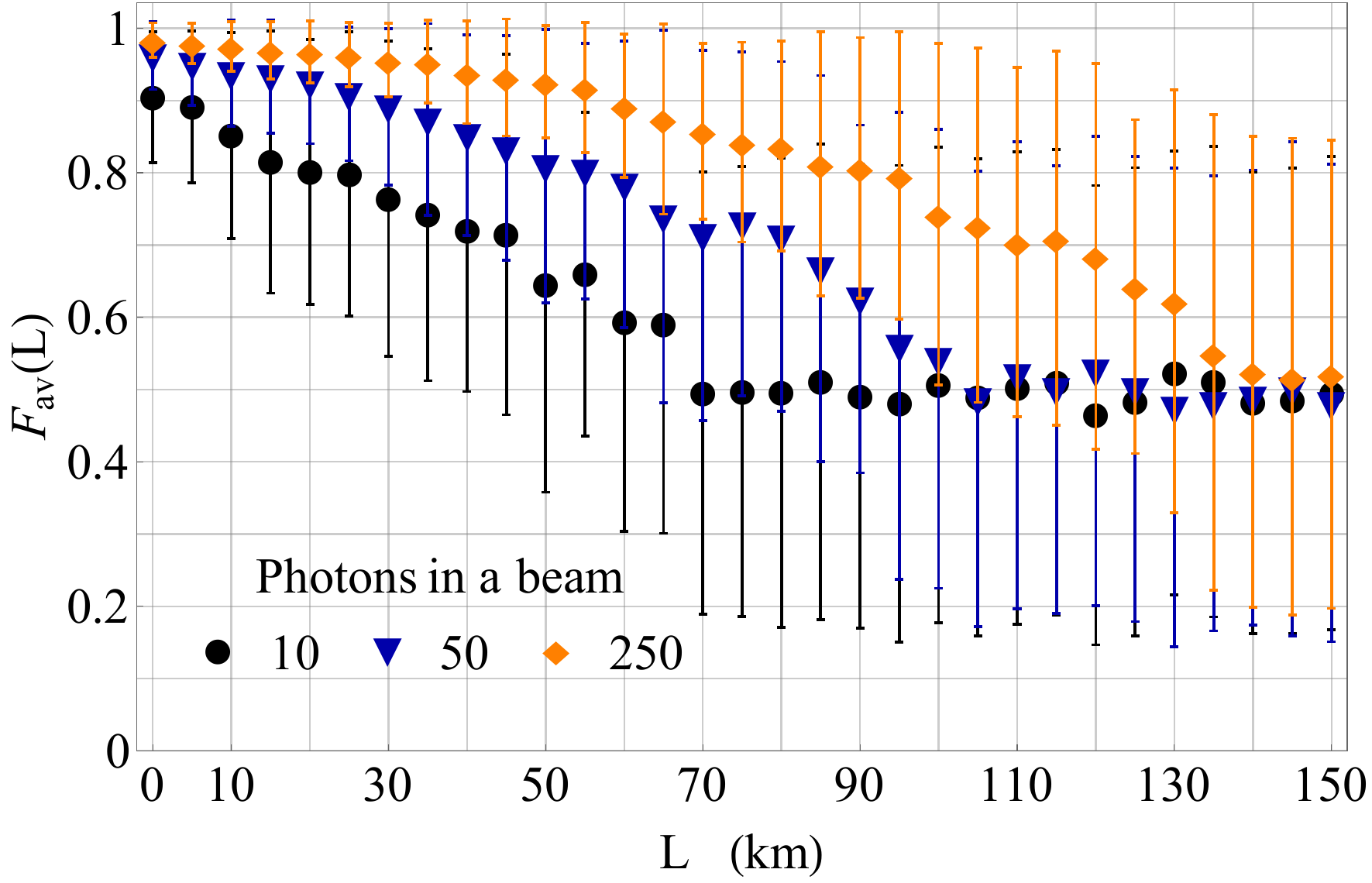}
\caption{Fidelity of qubit tomography, $F_{av} (L)$, for three numbers of photons per measurement. The attenuation coefficient is fixed: $\alpha = 0.2$ dB/km}
\label{qubitfidel1}
\end{figure}

First, we consider the efficiency of QST for single qubits. We selected a sample consisting of $220$ pure states that are distributed uniformly on the Bloch sphere. Then, each input state goes through the framework, and the average fidelity is computed. In \figref{qubitfidel1}, one finds the results presenting $F_{av} (L)$, for three numbers of photons per measurement. The results correspond to a fixed attenuation coefficient: $\alpha = 0.2$ dB/km.

Form \figref{qubitfidel1}, one can observe that the quality of state recovery degenerates as we increase the fiber. A longer fiber involves more attenuated photons, which reduces the precision of measurements due to the shot noise. If we compare $\mathcal{N}=50$ and $\mathcal{N}=250$, we notice that initially, both settings provided similar accuracy. However, as we increase the fiber, the setting with $250$ photons per measurement outperforms the other scenario. Also, one should notice that SD grows along with the fiber, which implies that the sample features more statistical dispersion. Finally, for $\mathcal{N}=10$, we see that even at the beginning, one is not able to properly estimate the state. It results from the impact of shot noise, which affects the smallest ensemble more significantly, even in the absence of fiber attenuation. Moreover, for $10$ photons per measurement, the results are more scattered, which means we cannot predict the efficiency for a particular state.

\begin{figure}[h]
\centering
\includegraphics[width=\linewidth]{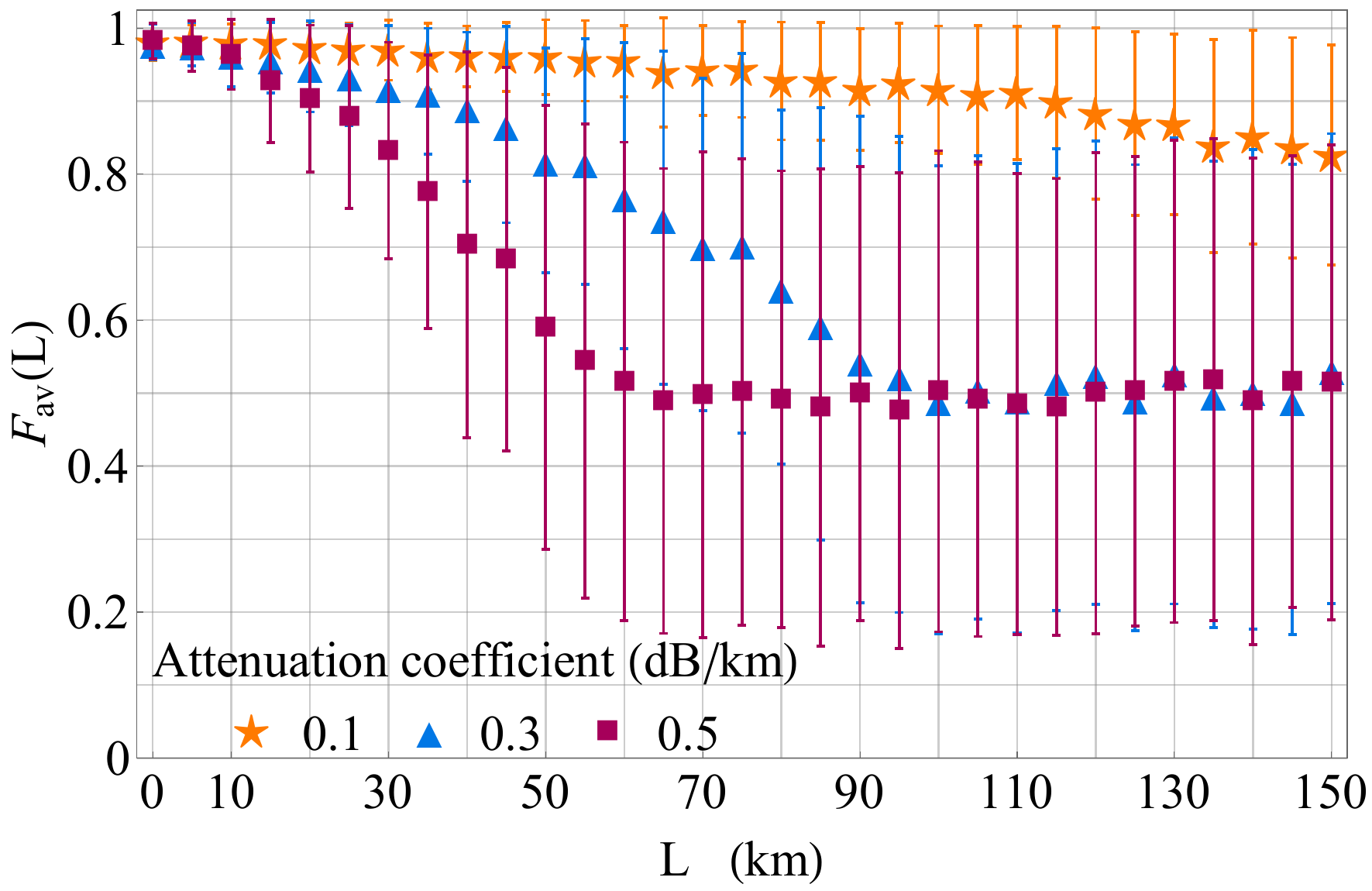}
\caption{Fidelity of qubit tomography, $F_{av} (L)$, for three attenuation coefficients. The number of photons per measurement is fixed: $\mathcal{N}=200.$}
\label{qubitfidel2}
\end{figure}

It is worth stressing that ultimately all plots converge to the value $F_{av} (L) \approx 0.5$ accompanied by a considerable SD. This tendency confirms the observations from the theoretical model introduced in Sec.~\ref{theorframework}. At some point, the photon loss is extremely significant, which, combined with the shot noise, makes the state estimation impossible. The QST framework results in random states and, consequently, a great value of SD. The average value of fidelity is close to $1/2$, which corresponds to quantum fidelity between any pure state and a maximally mixed state. 

Next, we consider $F_{av} (L)$ for three different attenuation coefficients while the number of photons is fixed: $\mathcal{N} = 200$. The findings are presented in \figref{qubitfidel2}. For $\alpha = 0.1$ dB/km, we witness stable precision of state estimation since the function decreases slowly. The results demonstrate that in this scenario, the framework is efficient even for longer fibers.

Then, for $\alpha = 0.3$ dB/km, we observe that the fidelity declines as we increase the fiber up to approx. $L=100$ km, when it stabilizes. There are some irregularities in the plot, which can be attributed to the randomness of noise that affects the measurements.

Lastly, if $\alpha = 0.5$ dB/km, the fidelity reduces rapidly. We notice a similar tendency as for $\alpha = 0.5$ dB/km, but the plot declines more sharply. For $L=60$ km, we obtain $F_{av} (60\:\mathrm{km}) = 0.52 \pm 0.33$, and then its value remains more or less constant. This observation suggests that one cannot efficiently estimate a qubit state in such conditions.

\subsection{Entangled qubits}\label{entqubits}

In this part, we apply the framework to one class of entangled qubits. More specifically, we investigate a following family of two-qubit states:
\begin{equation}\label{eq7}
    \ket{\Phi (\phi)} = \frac{1}{\sqrt{2}} \left(\ket{0}\otimes\ket{0} + e^{\phi i } \ket{1}\otimes\ket{1} \right),
\end{equation}
where $\{\ket{0},\ket{1}\}$ represents the standard basis in the two-dimensional Hilbert space and $0\leq \phi < 2 \pi$. We select a sample of $100$ states of the form \eqref{eq7} such that the relative phase $\phi$ covers the full range.

We focus on this particular class of two-qubit entanglement since it comprises the celebrated Bell states, i.e., $\ket{\Phi^+}$ and $\ket{\Phi^-}$, which are famous for multiple applications in quantum information and computation. In quantum optics, such kind of two-photon entangled states can be produced by spontaneous four-wave mixing (SFWM) in a dispersion-shifted fiber \cite{Takesue2009}, or spontaneous parametric down-conversion (SPDC) \cite{Marcikic2004}, and by a source that utilizes quantum dots \cite{Jayakumar2014}.

\begin{figure}[h]
\centering
\includegraphics[width=\linewidth]{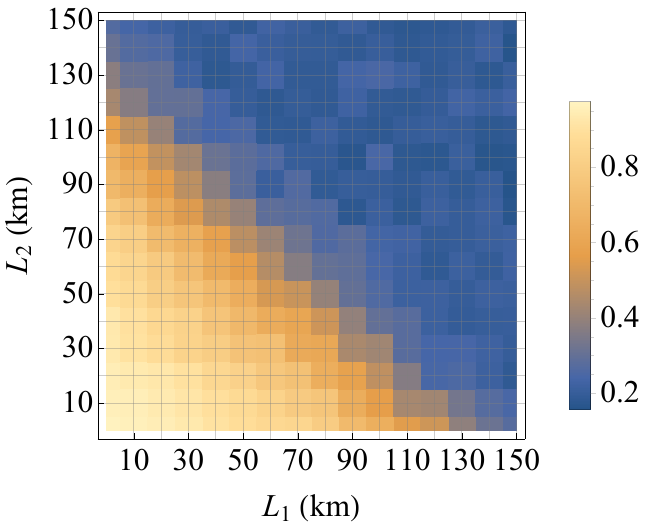}
\caption{Concurrence of entangled qubits tomography, $C_{av} (L_{1},L_{2})$. The number of photon pairs per measurement is fixed: $\mathcal{N} = 1\,000$ and attenuation coefficient is $\alpha = 0.2$ dB/km.}
\label{equbitcon}
\end{figure}

 The results with the fidelity of state estimation can be found in Appendix~\Ref{estimationfidelity}. Here, we concentrate on entanglement quantification because it is a crucial question in the context of practical applications. In \figref{equbitcon}, we present the average concurrence, $C_{av} (L_{1},L_{2})$, which corresponds to the estimated states. Originally, all input states were maximally entangled with concurrence equal to one. However, as a consequence of photon loss and shot noise, the measured states feature less entanglement. The plot \figref{equbitcon} allows one to track how the quantity of entanglement detected by the measurement scheme depends on fiber lengths.

For practical reasons, we are usually interested in detecting such a quantity of entanglement that is sufficient to announce the violation of the Bell-CHSH inequality \cite{Bell1964,Clauser1969}. Based on the concurrence, we can say that a quantum state $\rho$ allows of such a violation if $C[\rho] > 1/\sqrt{2}$ \cite{Verstraete2002,Hu2012}. From \figref{equbitcon}, one can conclude that $C_{av} (L_{1},L_{2}) >  1/\sqrt{2}$ as long as $L_1 + L_2 < 110$ km. This observation provides us with a threshold for detecting quantum correlations. By following the framework introduced in this work, one can determine an analogous criterion for a different number of photon pairs or distinct attenuation coefficients.

\subsection{Application: security of quantum communication}

We propose to implement the tomographic scheme in the context of security in quantum communication. When two parties share an entangled state, they can establish a private key by exploiting the quantum correlations associated with measurements \cite{Ekert1991}. However, the task is more challenging if we assume that an eavesdropper may hack into the communication system and impose some measurements on the photons. There are numerous works on the impact of an eavesdropper on a quantum cryptography system, see, e.g., Ref.~\cite{Acin2006,Gerhardt2011}. Usually, eavesdropper detection requires discarding some percentage of bits to compare them over a public channel. In this work, we introduce a tomographic approach to detecting an eavesdropper. From Sec.~\ref{entqubits}, we know how well entanglement can be retrieved in the presence of fiber attenuation. Thus, the key idea is that by performing entanglement quantification, we can discover a security threat if the result we obtain is below the theoretically estimated boundary.

As for the eavesdropper (Eve), we assume that she can plug in between the source and one party. She tries to avoid being caught easily, so she captures only a portion of photons in each beam, denoted by $p$ (where $0\leq p \leq 1$). Moreover, Eve performs the SIC-POVM measurements on her photons and lets them through the rest of the fiber. However, we do not know which operator is measured when. Thus, we consider her measurements to be random. Then, the post-measurement state of a photon pair is given by:
\begin{equation}
    \rho_{Eve}^k := (\mathbb{I}_2 \otimes M_k)  \ket{\Phi (\phi)} \!  \bra{\Phi (\phi)} (\mathbb{I}_2 \otimes M_k),
\end{equation}
where for each photon beam, $M_k$ is assigned randomly from the set $\{M_1,M_2,M_3,M_4\}$ which denotes the single-qubit SIC-POVM. As a results of the eavesdropping, the quantum state can be described by a statistical mixture
\begin{equation}\label{securitystate}
    \widetilde{\rho}_k = (1-p)  \ket{\Phi (\phi)} \!  \bra{\Phi (\phi)} + p  \,\rho_{Eve}^k.
\end{equation}

The state \eqref{securitystate} undergoes the series of $16$ measurements by the two-qubit operators, as described in Sec.~\ref{methodsection}. For each measurement, the Eve's operator $M_k$ is selected randomly to simulate the effect of eavesdropping.

\begin{figure}[h]
\centering
\includegraphics[width=\linewidth]{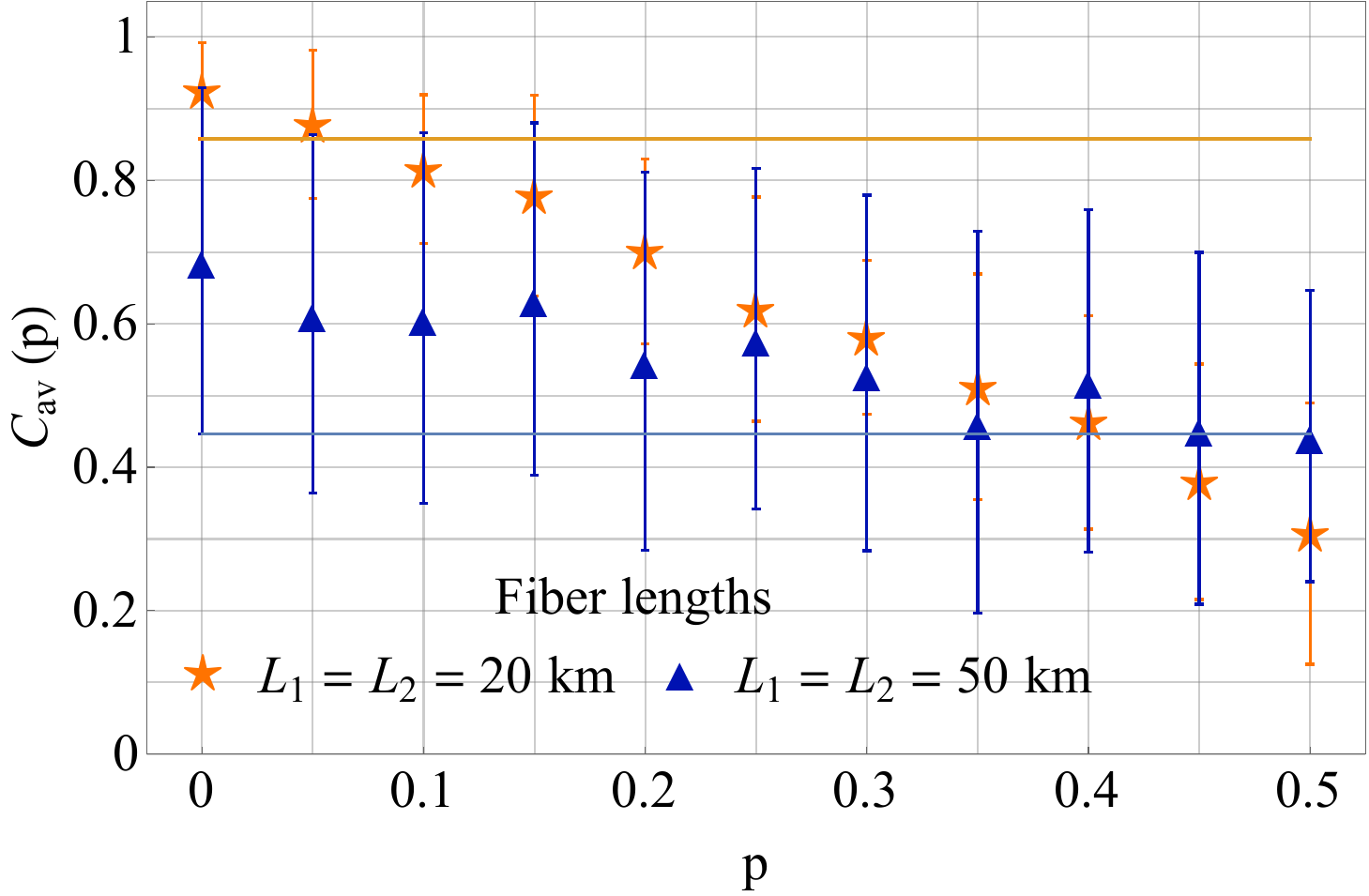}
\caption{Concurrence of a two-qubit state via quantum tomography affected by an eavesdropper. Error margins correspond to SD. The horizontal lines represent the lower bounds in the scenario with $p=0$.}
\label{securityfig}
\end{figure}

In \figref{securityfig}, we present the results of numerical simulations, i.e., the concurrence as a function of $p$, which is denoted by $C_{av} (p)$. The same sample of $\ket{\Phi (\phi)}$ was used as in Sec.~\ref{entqubits}. We assumed that the source emitted $1\,000$ photon pairs per measurement which are attenuated by the fiber. The eavesdropper was located in the middle of the fiber between the source and one of the parties. We considered two combinations of fiber lengths to discuss the efficiency of the eavesdropper detection.

First, we observe that $C_{av} (p)$ decreases monotonically when $L_1 = L_2 = 20$ km. When Eve is not active, we obtain $C_{av} (0) = 0.94\pm 0.06$, which corresponds to the results presented in \figref{equbitcon}. The lower bound of this interval is used as a frame of reference to determine a condition that allows one to detect the eavesdropper. We notice that for all $p\geq 0.2$, the entire intervals of $C_{av} (p)$ lie below the threshold. Based on this observation, we conclude that under such assumptions, one can detect the eavesdropper if Eve captures at least $20\%$ of photons.

On the other hand, if we take longer fibers such that $L_1 = L_2 = 50$ km, we obtain more statistical dispersion since $C_{av} (0) = 0.69\pm 0.24$. Due to photon loss, which is more significant in this case, the results for the sample are considerably affected by the shot noise and, as a result, more scattered. Again, the lower error margin is treated as a threshold for the eavesdropper detection. As we increase $p$, we see that $C_{av} (p)$ behaves non-monotonically. Although the average concurrence tends to decrease, we cannot detect the eavesdropper within the investigated range of $p$. In no case, the interval for the concurrence lies below the threshold.

In our example, we see that the fiber length can be considered a key limiting factor since the scenario is focused on the impact of fiber attenuation. This approach is in agreement with other quantum information protocols that investigate the efficiency of information exchange by photons transmitted through a lossy fiber. A remedy for the effect of photon loss would involve increasing the number of photon pairs per measurement. As we already know, a longer fiber can be compensated for by a larger ensemble of quantum systems, which could be realized by a longer-lasting light pulse. In conclusion, due to the interdependence between the parameters, for a given setting, one would need to determine the optimal duration of the light pulse to utilize the method proposed in this work.

\section{Qutrit tomography}

\subsection{Single qutrits}

\begin{figure}[h]
\centering
\includegraphics[width=\linewidth]{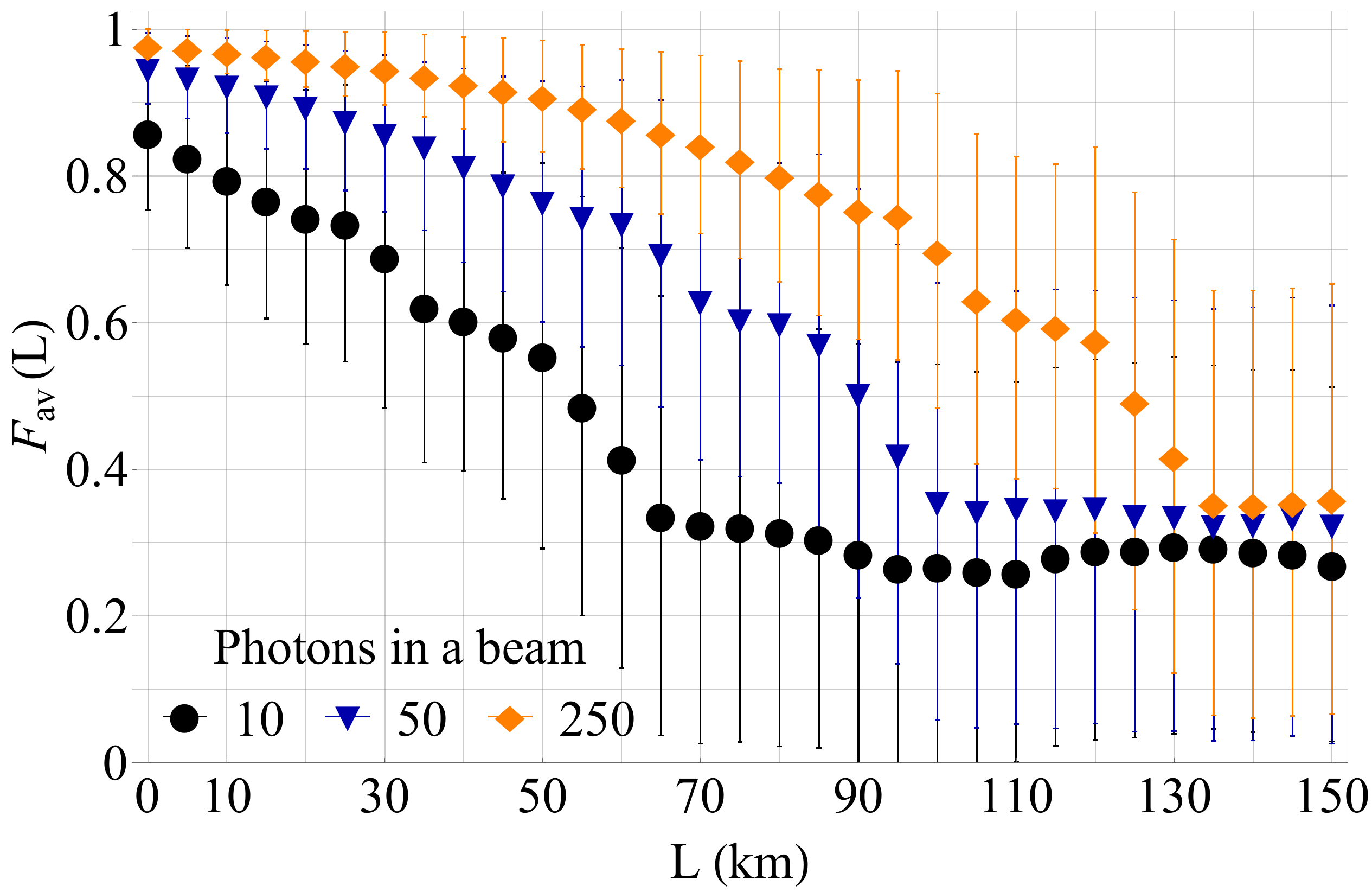}
\caption{Fidelity of qutrit tomography, $F_{av} (L)$, for three numbers of photons per measurement. The attenuation coefficient is fixed: $\alpha = 0.2$ dB/km.}
\label{fig:single_qutrit}
\end{figure}

In addition, we examine the performance of the QST framework with single qutrits. We took a sample of $5184$ pure qutrit states, which are equidistant across all parameters of the general representation of a qutrit pure state \cite{Szlachetka2021}. Each input state sequentially passes through the framework, and the average fidelity is calculated. In \figref{fig:single_qutrit}, we present the results of the average fidelity calculation, $F_{av} (L)$, for three numbers of photons per measurement: $\mathcal{N} = 10$, $50$, and $250$. The results are estimated for a fixed attenuation coefficient: $\alpha = 0.2$ dB/km.

The results presented in \figref{fig:single_qutrit} demonstrate that if we increase the fiber length, one can observe that the average fidelity in every case, $F_{av} (L)$, drops down nearly linearly in some range and then achieves a constant value roughly equal to the $1/3$. Deformation of the estimated state is caused by the absorption, which decreases the number of photons in measurements. Also, these findings are in line with the theoretical analysis presented in \figref{purity-entropy}. A tomography scheme can extract the reliable state when there is a sufficient probability of getting enough photons to extract the parameters that characterize the density matrix. This can be related to the entropy of the Fock state that declines as we increase $L$, which implies that the Fock state converges to the $\ket{0}$.

\begin{figure}[h]
\centering
\includegraphics[width=\linewidth]{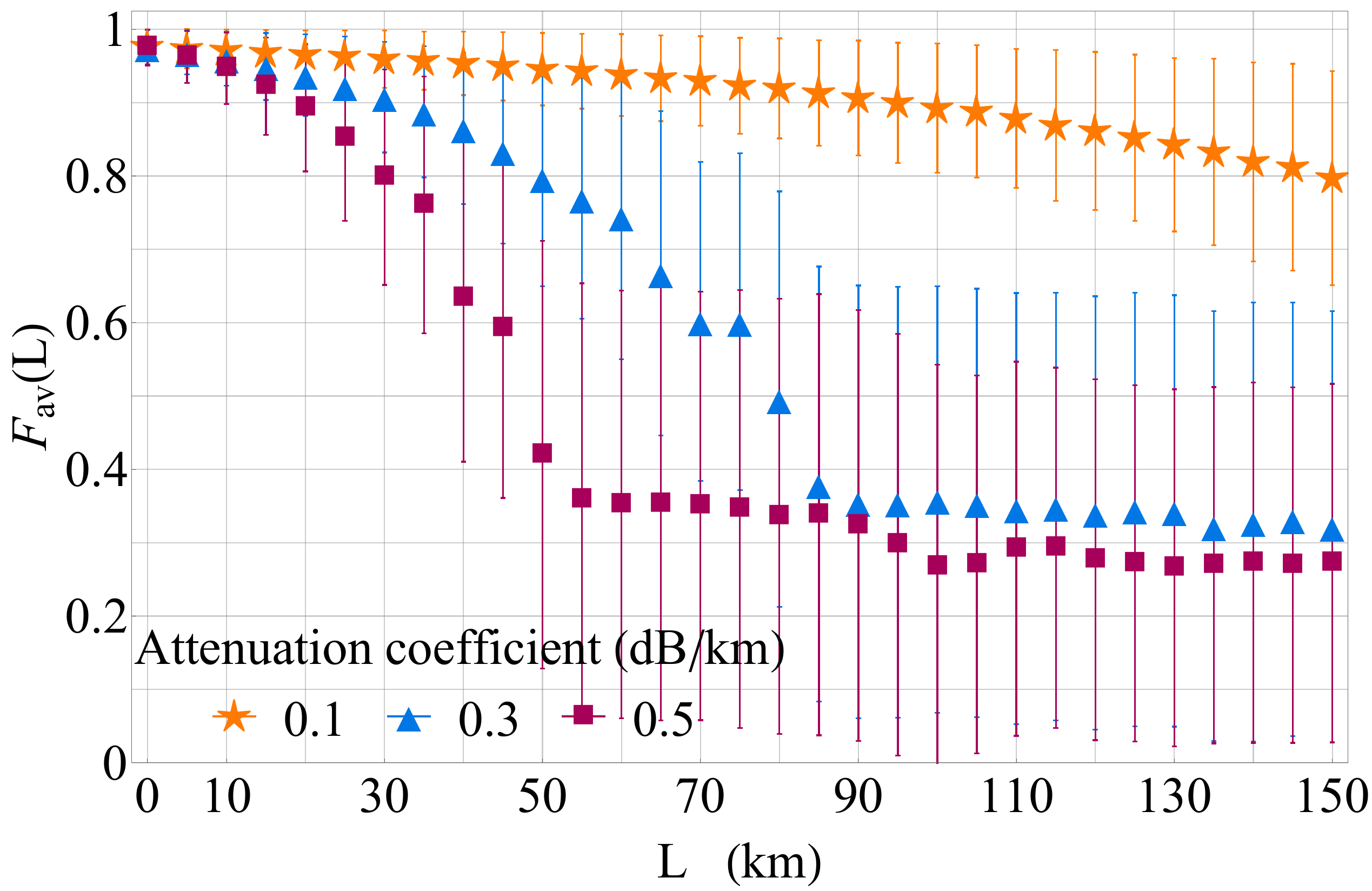}
\caption{Fidelity of qutrit tomography, $F_{av} (L)$, for three attenuation coefficients. The number of photons per measurement is fixed: $\mathcal{N}=200.$}
\label{fig:single_JS}
\end{figure}

Comparing these findings with qubits, we observe a faster degradation of the estimated state while increasing the length of the fiber. It can be concluded that one needs more measurements to achieve decent fidelity for a higher dimensional state. It stems from the need to estimate more parameters. Furthermore, different constant values of the average fidelity can be observed in the long fiber regime. In the qubit scenario, $F_{av} (L)$ approaches $1/2$, whereas, for qutrits, it converges to $1/3$. These values are connected to the scenario when we do not have any information about the measured state since the framework randomly assigns a density matrix to the data due to noise.

Increasing the number of input photons extends the length of the fiber for which we can achieve a reliable state. Thanks to this, we can predict, for a specific length of the fiber, the number of photons that we must be measured to get sufficient quality of estimation. Moreover, we can observe that the standard deviation increases with the length of the fiber, which is related to the statistical dispersion of the sample.

Then, we considered the properties of $F_{av}(L)$ for a constant number of photons: $\mathcal{N} = 200$, and three different attenuation parameters: $\alpha  = 0.1$, $0.3$ and $0.5$ dB/km. We observe faster degradation of state estimation for an increased damping factor. The trend of the decrease of the average fidelity is close to the linear for $0.1$ dB/km. This behavior is no longer valid for the highest attenuation coefficient $\alpha=0.3$ and $\alpha=0.5$ dB/km. First, we witness a rapid decline in the accuracy of state estimation. Later, for the attenuation coefficient $\alpha=0.3$ dB/km when $L\geq 85$ m and $\alpha=0.5$ dB/km when $L\geq 55$ m, the function $F_{av}(L)$ stabilizes and maintains roughly a constant value close to $1/3$. This feature is analogous to the characteristics of $F_{av}(L)$ observed for qubits for the same parameters, see \figref{qubitfidel2}. However, in the case of qutrits, the constant value of $F_{av}(L)$ is achieved for shorter fibers than in the case of qubits. This implies that a long fiber combined with a higher attenuation coefficient is highly detrimental to a qutrit state recovery.

\subsection{Entangled qutrits}\label{entanqutrits}

To study the performance of QST for entangled qutrit states, we chose a class of maximally entangled states:
\begin{equation}\label{eq8}
    \ket{\Theta (\phi)} = \frac{1}{\sqrt{3}} \left(e^{i \phi }\ket{0}\otimes\ket{2} + \ket{1}\otimes\ket{1} + e^{i \phi  }\ket{2}\otimes\ket{0} \right),
\end{equation}
where $ \{\ket{0}, \ket{1}, \ket{2}\}$ represents the standard basis in the three dimensional Hilbert space and 0 $ \leq  \phi< 2\pi$. The sample consists of $100$ states equally spaced due to the relative phase $\phi$. We selected the input states $\ket{\Theta (\phi)}$ since such systems can be easily produced with a spatial degree of freedom in the SPDC process \cite{Pimenta2013}.

\begin{figure}[h]
\centering
\includegraphics[width=\linewidth]{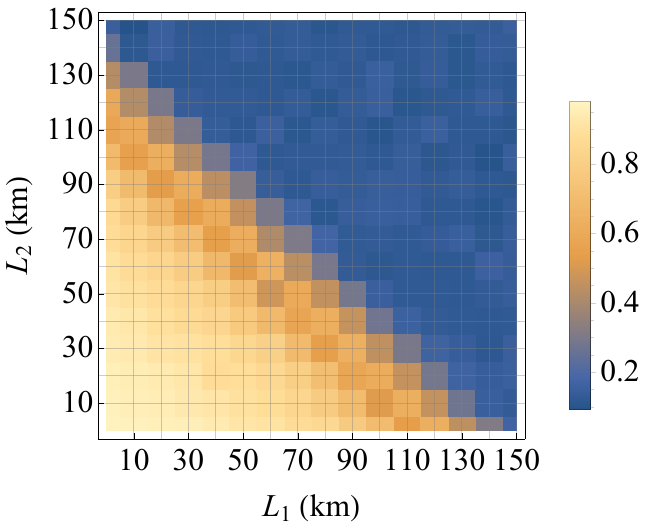}
\caption{Negativity of entangled qutrits tomography, $\mathrm{N}_{av} (L_{1},L_{2})$. The number of photon pairs per measurement is fixed: $\mathcal{N} = 1000$ and the attenuation coefficient is $\alpha = 0.2$ dB/km}
\label{fig:etanqutritetan}
\end{figure}

In \figref{fig:etanqutritetan}, we present the mean negativity, which represents the amount of entanglement of the reconstructed two-qutrit state after the transmission in fibers in two arms. The simulations were performed for $\mathcal{N} = 1000$ and attenuation coefficient $0.2$ dB/km. The amount of entanglement tends to decrease monotonically as the optical fiber length becomes greater in both directions. The measurement of negativity gives a high amount of entanglement when one party has a long fiber and the other uses a short channel. Due to the necessity to estimate many parameters, the performance for entangled qutrits is more distorted by the absorption in optical fibers than in the qubit case. Moreover, the negativity measurement for the employed sample of states with deformation stemming from the absorption in fiber is well-behaving. Minor inaccuracies that occurred in \figref{fig:etanqutritetan} (e.g., the function $\mathrm{N}_{av} (L_{1},L_{2})$ is not always monotone) can be attributed to the random nature of noise. Both photon loss and shot noise are governed by probability distributions, which results in random fluctuations of the figures of merit.

\section{Conclusions and outlook}

In this paper, we have introduced the formalism of photon loss within the theory of open quantum systems. The binomial distribution and the Beer-Lambert law were implemented to describe how the Fock state changes when a beam of photons travels through a fiber. The impact of fiber attenuation can be modeled by amplitude-damping quantum channels, which is in line with other quantum phenomena such as spontaneous emission. In addition, the evolution of the Fock state in the domain of the fiber length can be represented by a master equation with a linear generator. Both theoretical frameworks provide valuable insight into the process of photon loss. Also, the framework appears applicable in generating random numbers. 

In practice, polarization-entangled photon sources suffer from limited performance \cite{Montaut2017,Meyer-Scott2018}, which means that produced photon pairs should be distributed efficiently. Therefore, the scheme was adapted to qubit and qutrit tomography, while special attention was paid to entangled states. With selected figures of merit, we could track both the precision of state reconstruction and the quantity of entanglement detected by the framework. The results allow one to observe how the quality of transmission degenerates as we increase the fiber length. The findings of the paper may have relevant implications for future experiments. One needs to properly select the duration of a light pulse to guarantee a sufficient number of photons per measurement. In other words, the awareness of how fiber attenuation impacts the transmission of photons will help an experimenter to adjust the input power of the source so that the detectors receive a satisfactory signal.

Furthermore, we demonstrated that the framework could be implemented to detect an eavesdropper in quantum communication based on entangled photon pairs. If a sufficient percentage of photons is disturbed by the eavesdropper's measurements, we can expose the eavesdropper by comparing the obtained concurrence with a threshold value. This method can be an alternative to other approaches to the eavesdropper detection problem.

There are remaining open problems that can be addressed in forthcoming papers. Most of all, fluctuations of the source should be taken into account. Even if we tune the source to a specific power, the number of photons emitted in a single shot can vary. More specifically, it should not be treated as a constant value but rather as a random variable generated from a distribution that characterizes the source. Furthermore, apart from fiber attenuation, we should also consider scattering processes that occur during the transmission of photons and influence the detection. A thorough description of all factors that impact the photonic tomography will lead to a better understanding and, presumably, more effective practical implementations of quantum information protocols.

\section*{Acknowledgments}

A.~C. was supported by the National Science Centre in Poland, grant No. 2020/39/I/ST2/02922. J.~S. acknowledges financial support from the Foundation for Polish Science (FNP), project First Team co-financed by the European Union under the European Regional Development Fund, grant no. First Team/2017-3/20, and National Laboratory of Atomic, Molecular and Optical Physics.

\appendix

\section{Fidelity of entangled states estimation}
\label{estimationfidelity}

\begin{figure}[h!]
\centering
\includegraphics[width=\linewidth]{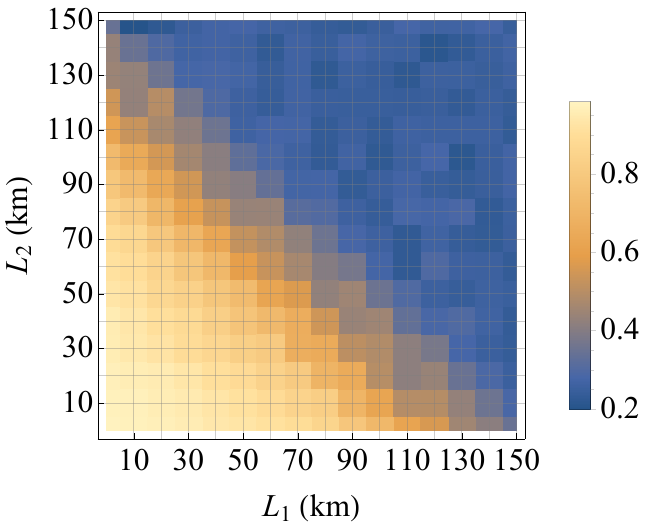}
\caption{Fidelity of entangled qubits tomography, $F_{av} (L_{1},L_{2})$. The number of photon pairs per measurement is fixed: $\mathcal{N}=1\,000$ and attenuation coefficient is $\alpha = 0.2$ dB/km.}
\label{equbitfidel}
\end{figure}

\begin{figure}[h!]
\centering
\includegraphics[width=\linewidth]{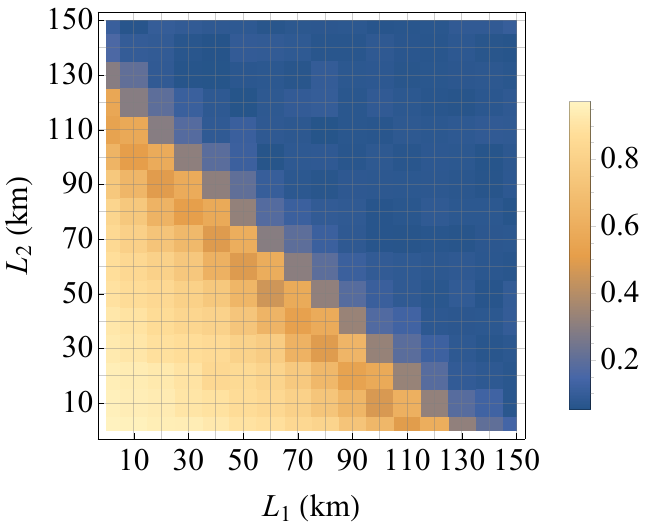}
\caption{Fidelity of entangled qutrits tomography, $F_{av} (L_{1},L_{2})$. The number of photons pairs per measurement is fixed: $\mathcal{N}= 1000$ and the attenuation coefficient is $\alpha = 0.2$ dB/km.}
\label{fig:fidelqutritentan}
\end{figure}

In the appendix, we provide plots presenting the fidelity of entangled states estimation.

In \figref{equbitfidel}, one finds the average fidelity, $F_{av} (L_{1},L_{2})$, corresponding to entangled qubits estimation, as discussed in Sec.~\ref{entqubits}. The map allows one to follow the precision of state reconstruction for different combinations of fiber lengths. We assumed that the number of photon pairs per measurement and the attenuation coefficient remained unchanged.

In \figref{fig:fidelqutritentan}, we display the average fidelity depending on lengths of two fibers, with $\mathcal{N} = 1000$ and the attenuation coefficient: $0.2$ dB/km, as discussed in Sec.~\ref{entanqutrits}. The state deformation becomes larger as the lengths of the optical fibers increase in both arms. However, it can be concluded that the region of reliable state estimation can be defined by the following constraint: $L_1 + L_2\leq 90$ m.

\end{document}